\documentclass[referee,usegraphicx]{mn2e}

\begin{document}           
\title[Optical polarization observations: NGC 6124]{ Optical polarization observations in the Scorpius region: NGC 6124\thanks{Based on observations obtained at  Complejo Astron\'omico El
Leoncito (CASLEO), operated under agreement between the CONICET and the
National Universities of La Plata, C\'ordoba, and San Juan, Argentina.}}

\author[Vergne et al.]{M. Marcela Vergne,$^{1,2,3}$ Carlos Feinstein,$^{1,2,3}$  
Ruben Mart\'{\i}nez,$^{1,3}$
\newauthor  Ana Mar\'{\i}a Orsatti$^{1,2,3}$ and Mar\'{\i}a Paula Alvarez$^{1}$\\
$^1$Facultad de Ciencias Astron\'omicas y Geof\'{\i}sicas, Observatorio Astron\'omico, Paseo del Bosque, 1900 La Plata, Argentina \\
$^2$Member of Carrera del Investigador Cient\'{\i}fico, CONICET,Argentina\\
$^3$Instituto de Astrof\'{\i}sica de La Plata, CONICET}
 
\maketitle


\begin{abstract}

We have obtained optical multicolour (UBVRI) linear polarimetric data for 46 of
the brightest stars in the area of the open cluster 
NGC 6124 in order to investigate the properties of the interstellar medium (ISM)
that lies along the line of sight toward  the cluster. Our data yield a mean
polarization efficiency of $P_V/E_{B-V}=$3.1$\pm$0.62, i.e., a value lower than
the polarization produced by the ISM with normal efficiency for an average colour excess
of $E_{B-V}=0.80$ as that found for NGC 6124. Besides, the polarization shows
an orientation of $\theta \sim $8$^\circ$.1 which is not parallel to the Galactic Disk,
an effect that we think may be caused by the Lupus Cloud.  Our analysis also indicates
that the observed visual extinction in NGC 6124 is caused by the presence
of three different absorption sheets located between the Sun and NGC 6124.
The values of the internal dispersion of the polarization ($\Delta P_V\sim 1.3\%$) and
of the colour excess 
($\Delta E_{B-V}\sim 0.29$ mag) for the members of NGC 6124 seem
to be compatible with the presence of an intra-cluster dust component.
Only six  stars exhibit some evidence of intrinsic
 polarization.
Our work also shows that polarimetry provides an excellent tool to distinguish
between member and non-member stars of a cluster.

\end{abstract}

\section{Introduction}

The open cluster NGC 6124 (l= 340.$^\circ$8, b=+6.$^\circ$1) is an intermediate-richness,
detached cluster of slight concentration with stars in a wide range of 
brightness located in the Scorpius region. This cluster is seen as projected on the edge
of a dust cloud which obscures some of its stars, on the north-western part of
cluster area. NGC 6124 contains several red giant stars.

This cluster has been studied photoelectrically by Koelbloed (1959), but he only made
rough estimates of the distance and of the age of the cluster. In order to get
a more accurate determination of those parameters, the study of Koelbloed was extended
photoelectrically and photographically to a fainter magnitude limit by Th\'e (1965).
These initial studies of the cluster revealed important discrepancies
between the values of the parameters that they obtained. Motivated by some
evidence that this cluster is affected by differential reddening, Pedreros (1987)
obtained a new set of UBV photoelectric observations for NGC 6124 and analyzed them
using a reddening-line slope adequate for this region of the sky.
The distance to NGC 6124 that he estimated using the less evolved stars is $563 \pm
10$ pc. He also found an age  of $1.0$ x $10^{8}$ yr and a mean colour
excess of $E_{B-V} = 0.80 \pm 0.06$. 

A study of the interstellar polarization of NGC 6124 is warranted for three
reasons: it provides information on the dust itself, it is a means to trace the
galactic magnetic field and it is also useful to establish cluster membership.
The comparison between the polarization and extinction values observed along
the same line of sight provides tests for extinction models and grain alignment.
The latter involves several processes acting simultaneously, but on different
time-scales. The rotating dust grains get a substantial magnetic moment that allows them
to precess rapidly about magnetic field lines and, as a result, the grains
conserve their orientation relative to that of the magnetic field when
the latter fluctuates, forcing the axes of the grains to be aligned with 
respect to the angular momentum of the grains (Lazarian and Cho 2005). Therefore,
the observed polarization vectors map the projected direction of the magnetic
field on the plane of the sky , and this allows the investigation of the
structure of both the macroscopic field in our Galaxy (Mathewson and Ford 1970; Axon and
Ellis 1976) and the local field associated with the individual clouds (Goodman et al. 1990).

  Open Clusters are very good candidates for polarimetric observations, because
many of them have been studied through photometric and spectroscopic techniques  and
detailed information on the colour, luminosity, spectral type, and other properties
of their stars is readily available. Thus, the physical parameters of the
cluster and its stars can be obtained, and adding polarimetric observations
we can study the distribution, size and efficiency of the
dust grains which polarize the starlight and the different directions of the
galactic magnetic field along the line of sight to the cluster.
As some of the open clusters spread over a significant area, it is possible to analyze the
evolution of the physical parameters of the dust over the region. Besides, it is
also possible to detect the presence (if any) of intra-cluster dust and ,with additional
observations of non-member stars in the same region, to investigate the interstellar
dust distribution along the line of sight to the cluster. Finally, the
polarimetric data can be used to detect the location of any energetic phenomenon that may have
occurred in the history of a cluster (Feinstein et al. 2003), and  very important
by-products of these studies are the detection of stars displaying polarization
of  non-interstellar origin ,such as stars with extended atmospheres (e.g., Be
stars), and of dust associated to possible binary systems, or surrounding the stars (due to
evolution or as a formation remnant).
 
 Here we present multicolour (UBVRI) measurements of linear polarization
vectors for stars observed in the direction to NGC 6124 and we use them to investigate
the properties of the dust located along the line of sight to  the cluster . In
the following sections, we will discuss the observations, the data calibrations and the results
obtained both for the individual stars and for the cluster as a whole.

\section{Observations and data reduction}

The linear optical polarimetry was obtained during several observing runs at the
Complejo Astron\'omico El Leoncito (CASLEO) in San Juan, Argentina, during 2003-2007. The
Torino five-channel photopolarimeter (Scaltritti et al. 1989, Scaltritti 1994) was used
on the 2.15-m telescope. All observations were obtained++ simultaneously through
the five Johnson-Cousins broad band {UBVRI} filters centered at $\lambda_{Ueff}=0.360\ \mu m$,
$\lambda_{Beff}=0.440\ \mu m$, $\lambda_{Veff}=0.530\ \mu m$, $\lambda_{Reff} =0.690\ \mu m$ and
$\lambda_{Ieff}=0.830\ \mu m$. Standard stars for null polarization and for the zero point of the
polarization position angle were observed several times each night for calibration purposes.
The instrumental polarization was obtained from the null polarization standards and subsequently
used to correct the values derived for the program stars. Polarization standards were used to
refer the observed polarization angles to the equatorial system.
Table 1 shows the average of the observations of the null polarization standards. Only in the last
run a correction for instrumental polarization (hereafter: IP) on the U filter was needed. For
the other filters the value of the IP was of the same order of magnitude of the
observational errors which means that the IP is either negligible, or so low that
it cannot be determined.   

The polarimetric observations are listed in Table 2 which shows, in self
 explanatory format, the stellar identification as given by Koelbloed (1959), 
the polarization percentage
 ($P_{\lambda}$) and the position angle of the electric vector
 (${\theta}_{\lambda}$) through each filter, as well as their respective
 mean errors computed considering the photon shot noise as the dominant source
 of errors (Maronna et al., 1992). Since the Torino
 photopolarimeter collects photons simultaneously in all the filters (UBVRI),
 the final data from each filter may be of different quality, especially those
 in the U band.  Therefore, observations whose values were below the 3$\sigma$
 level were ruled out and are not included in Table 2.
  Three stars (\#6, 9, and 31) were not rejected, and are included in Table 2
although their
measurements showed high errors, because they are evidently low polarization
 stars located near the Sun.

A total of 46 stars were polarimetrically observed in the area of NGC 6124.
We have 21 stars in common with the photoelectric observations of 
Pedreros (1987) and other 6 stars that were quoted as red giants
 by Koelbloed (1959). The membership information for each star, indicated in
 the first column of Table 2, was obtained from Pedreros (1987),
Th\'e (1965), Baumgard et al. (2000) and from the Webda (http://www.univie.ac.at/webda/).
Note, however, that not all the stars observed for the present work have membership information.


\section{Results}

The sky projection of the V-band polarization vectors for the observed stars
 in NGC 6124  are shown in Fig. 1. The dashed line superimposed to the
 figure is indicating the orientation of the projection of the Galactic Plane (GP), which has a P.A. of  
$\theta_{GP} \sim 44^{\circ}$.  Most of the 
 polarization vectors are not aligned with this direction.  This is not the usual finding, 
because dust particles in most aged clusters are expected to have enough time to relax and
 polarize the light in the same orientation as the GP in the region (Axon et al.,1976).
Normally this is a sign that a perturbation has happened to the dust (Ellis et al., 1978).

Fig. 2 shows the relation that exists between $P_V$ and $\theta_V$ 
 for members (full points), red giants (starred symbols),
 non-members (open triangles) and stars without any membership data (open circles). We can see
 a high scatter in angles, probably as a
 consequence of the presence of intra-cluster dust or a patchy structure of dust clouds on line of view to the cluster.
Also, we observe a spread in polarization values of $\Delta P_V=1.3\% $ in the
 stars historically considered as members
of the cluster (excluding stars 3 and 32 with
low polarization values, whose memberships will be analyzed in the next
sections). 

Fig. 3 shows the polarization angle distribution. The majority of the stars
have their angles in the range of $-10^{\circ} < \theta_V < 20^{\circ}$ (grey bars)
 and
 the angle distribution of the polarization
vectors of the members (dark bars) has
 a similar behaviour to that of the total sample, with a
$\Delta \theta_V > 20^{\circ}$ (with a $FWHM\sim8^{\circ}.5$). The mean
orientation of the polarization vectors in the cluster is $\bar{\theta}_V\sim8^{\circ}.1$
(vertical dashed line) but the orientation of the GP in
 the region is $\theta_{GP}\sim44^{\circ}$ (vertical solid line). So both
 directions differ in $\sim36^{\circ}$ degrees.
 All the stars that have $\theta_V > 25^{\circ}$ are already known non-members.

  A group of 7 red giant stars (\#1,14,29,33,35,36, and 41) is included in our
sample, 6 of them photoelectrically observed by
Koelbloed (1959). According to his data, they are K and M giants and
members of the cluster. Pedreros (1987) adopted the photometric
data from Koelbloed (1959) and their locations in his colour-colour diagram show
that these stars would belong to the cluster. These
evolved stars have extended atmospheres, therefore they are candidates
to have a non-interstellar component in their polarizations.
The polarization value, in the visualfilter, of each of the red giant
 stars is similar (see Table 2),
 excepting stars \#1 and 41 that 
have extreme polarization values ($P_V$=3.75\% and $P_V$=1.58\%, respectively).
The low polarization of the star \#41 suggests that it could be
 depolarized due to the presence
 of an intrinsic component.
 The polarization angles of the giant stars are lower
 than 10$^{o}$,
with the exception of stars \#33 and 36. The first one is classified
as M0III in the literature, and according to Th\'e (1965) it can probably be
a distant giant M star not belonging to the cluster. The second star (\#36) has
several spectral types in the literature (G2III, G8III, K2III), but according
to the Washington Visual double catalog (Worley 1996), it has a companion
with magnitude 14.20 mag. The binary nature of this star could explain
 the variation of the polarization
 angle with the wavelength.

\section{Analysis and discussion}
\subsection{Serkowski's Law}

To analyze the data, the polarimetric observations in the five filters were fitted in each star of our sample using Serkowski's 
law of interstellar polarization (Serkowski 1973). That is:

$P_{\lambda}/P_{\lambda max}=e^{-Kln^2(\lambda_{max}/\lambda)} \  \  \ \ [1]$

We will assume that, if polarization is produced by aligned interstellar
 dust grains, 
the observed data (in terms of wavelength, UBVRI) will then follow [1] and each star will have 
a $P_{\lambda_{max}}$ and $\lambda_{max}$ values.

To perform the fitting we adopted $K=1.66 \lambda_{max}+ 0.01$ (Whittet et al.
 1992). For all stars in the sample we computed the $\sigma_{1}$ parameter (the unit
 weight error of the fit) in order to quantify the departure of our data from
 the ``theoretical curve'' of Serkowski's law. In our scheme, when a star shows
$\sigma_{1} > 1.5$, it is indicating the presence of intrinsic stellar
 polarization. Also, $\lambda_{max}$ values can be used to test
 the origin of the
 polarization: those stars which have $\lambda_{max}$ much lower than the
 average value of the interstellar medium (0.545 $\mu m$, Serkowski, Mathewson
 and  Ford 1975) are candidates to have an intrinsic component of polarization
 as well. Another criterion to detect
 intrinsic stellar polarization comes from computing the dispersion of the
 position angle for each star normalized by the average of the position angle
 errors ($\bar{\epsilon}$). The values obtained for  $P_{\lambda_{max}}$, the
 $\sigma_{1}$ parameter, $\lambda_{max}$, and $\bar{\epsilon}$ 
together with the identification of stars, are listed in Table 3. The expression used to calculate the $\sigma_{1}$ parameter is also shown in the footnote.
No star in Table 3 has its $\lambda_{max}$ much lower than the average for the ISM.


    Fig. 4 displays $P_{\lambda}$ and $\theta_{\lambda}$
plots for six stars (\#13,26,27,36,42, and 46) with probable indications of intrinsic polarization.
The presence of a non-interstellar component in the measured polarization
 of the light from a star causes a
mismatch between observations and the Serkowski's curve fit, and/or a rotation
in the position angle of the polarization vector. This mismatch is
clearly shown in the $\sigma_{1}$ value  for star \#42 ($\sigma_{1}$= 1.81), and its 
variations in  $P_{\lambda}$ is a plane curve, indicating more than one component in polarization.
In the other five stars,  the presence of an intrinsic
component of polarization is seen through the variation of $\theta_ {\lambda}$
($\bar{\epsilon}$ parameter). In particular, the $P_U$ of star \#36 is out of the Serkowski's fit and, as it was
mentioned before, it is a known binary system.


\subsection{Stokes plane and memberships}

        To analyze the characteristics of the interstellar medium in the region
 of NGC 6124, we plot the normalized individual Stokes parameters in the
visual filter of the polarization
vector {\bf $\vec P_V$}, given by $Q_V=P_V cos(2 \theta_V)$ and $U_V=P_V sin(2 \theta_V)$,
 which represent the vector's equatorial components, for each of the observed 
stars. In this figure the stars are plotted according to the
 literature with their
 initial membership status. Filled circles are used for members, open triangles
are used for non-members, starred symbols are used for red giant stars and open
 circles are used for those stars without membership information.

      The polarimetric technique can help to solve membership problems. This
 type of plot ($Q_V$ vs. $U_V$) used in combination with photometric information
is useful to separate between members and non-members.

        In Fig. 5 the point of coordinates $Q_V=0$ and $U_V=0$ indicates the
 dustless solar neighborhood, and any other point represents the direction
of the polarization vector with modulus $P_V=\sqrt{Q_V^2+U_V^2}$ as seen
 from the
Sun. Again, this figure shows a high dispersion
in angles.

\subsubsection{Stars close to the Sun}
         On the left side of the diagram, near point (0,0),
 there is a group of three stars (\#6,9,and 31)
 with very low polarization values
 (0.11\%, 0.23\% and 0.07\%, respectively). Star 6 was observed by Pedreros
 (1987) and classified as a non-member of spectral type G5. Star 9, 
according to Th\'e (1965), may be a non-member variable star, and star 31 is of
spectral type F8V (Houk et al. 1975). The low polarization of these stars
is compatible with their small reddening. Therefore, 
 they could be confirmed polarimetrically as non-member stars.

 Stars \#3 and 32 (originally considered as members) have
 their polarization values of 0.89 \% and 1.02 \%, respectively. 
 From its photometric data, star 3 (of spectral type G2V)  
is located near the
Sun ($\sim$ 100 pc), with an $E_{B-V}=0.08$ mag. compatible with its 
polarization. In Fig. 1 this star is projected
on the north-west side of the sky, where a big dust cloud is
observed, obscuring several members on this part of the cluster.
 Therefore, if it were a member, its light would be expected to be highly
 polarized, but it is not. For Star \#32,
according to its Q parameter (Schmidt-Kaler 
1982), we obtain a F0-4V approximate spectral type and a distance of
$\sim$160 pc. From these data, the polarization of the light of both stars
 is likely caused in a dust cloud near the Sun, and for that we propose them
 as non-member stars.
 
\subsubsection{Non-member stars}

 In Fig. 5 there are two groups of non-member stars with different 
orientations of polarization vectors, but with similar values in polarization.
  There is a group made up of 4 non-member stars (\#34,39,44 and 45) with
mean values $\bar{\theta}_V$=17$^{o}$ and $\bar{P}_V$=1.98
 \%.
   Koelbloed (1959) suggested that \#44 and \#45 were members of the cluster
with a common spectral type AO, but their reddenings are smaller than those
of other A type stars in the main sequence of the cluster; and besides, \#44
is a B8II star according to Houk et al. (1975). 
 Also, if we take into
 account that these four stars are very close in projection
 on the sky (on the west size, see Fig. 1) and in the $Q_V$-$U_V$ plane,
 indicating similar polarimetric characteristics, probably they 
are polarized by the same dust cloud. From their photometric data, they are
 situated nearby the cluster ($\sim$ 400 pc from the Sun) and in front of it.
Therefore, we may confirm them polarimetrically as foreground non-member stars.
 Also, stars \#36 (red giant) and \#42 (member) could form part of this group,
 but they show some evidence of a non-interstellar component. 

Other two non-member stars, \#20 and 28, have very different polarimetric
 orientations
 (2$^{o}$.7 and 0$^{\circ}$.8, respectively) in comparison to the preceding
non-members. Both stars are projected
 on the central part of the cluster (Fig. 1), and to a mean distance of the Sun
of approximately 200 pc (from their photometric data). 
 The scatter between the two non-member groups in their position angles 
 ($\Delta\theta_V \sim 16^{\circ}$) may be showing
 that the light from these stars is polarized by several overlapped dust
 components along the line of sight to the cluster, with different
 orientations of the local magnetic fields.

\subsubsection{Cluster stars}

 Most stars in our sample have their $Q_V$ values in the range (1.29, 2.6),
 with a high scatter in polarization angles.
 In this range we observe a group of stars showing some degree of concentration,
considered most of them as members in the literature.
 To derive mean values,
 in angle and polarization for the cluster, we use seven member
stars (\#8,18,21,30,37,38, and 40), obtaining $\bar{P}_V$=2.11\% and
 $\bar{\theta}_V$=8$^{\circ}$.2. The wavelength of maximum polarization
for the same group of stars amounts to $\bar\lambda_{max}=0.57\pm0.04 \mu$m, a
 value very close to that of the ISM.
 According to their locations on the $Q_V$-$U_V$ plane
stars \#22 and 26 (originally non-members) may be polarimetrically
 considered now as members.
 
  On the upper side of this group, we find three stars \#5,33, and 43
 with 
 polarization values higher than the mean value for the cluster
 ($P_V$=2.77\%, 2.78\%, and 2.7\%, respectively). Star \#5, of spectral type 
B6II/III (Houk et al 1975),
 is considered as a member by Pedreros (1987), with a 60\% of
  membership probability
 (Baumgardt et al. 2000). Polarimetrically, 
it shows characteristics in angle similar to the members of the cluster,
 in spite of its 
high polarization value. Its estimated distance, according to the
spectral type, is approximately of 570 pc.
As mentioned before, star \#33 has a spectral type M0III and probably it is a
distant giant not belonging to the cluster. Star \#43 (spectral type B2III-IV)
is considered a member with a membership probability of 77\% (Baumgardt et al.
2000), but the orientation of the
polarization vector (${\theta_V}$=18$^{o}$.1) is very different from the
mean value of the cluster. As this star is an evolved object, an intrinsic
polarization component may be expected, but our indicators do not show it.
Therefore, these three stars may be possible non-members, due to their position
in the $Q_V$-$U_V$ plane.

 Stars \#11, 12, 13 and 16 are located on the lower side of
 the $Q_V$-$U_V$ plane.
 Star \#16 is considered as a member by Pedreros (1987), but its polarization
 value (1.63\%) and orientation (${\theta_V}$=176$^{o}$) are both lower than
the corresponding values of the rest of members. The estimated spectral
type for this star, based on photometric data, gives A8V with a
 $E_{B-V}$=0.39 mag. and a distance to the Sun smaller than that of the cluster. In
Fig. 1, it lies near stars \#20, and \#28 on the central part of the cluster,
 and also in the $Q_V$-$U_V$ plane,
 which indicates similar polarimetric characteristics in the dust
causing the polarization. Polarimetrically, we conclude that star \#16 may be a
non-member in front of NGC 6124. Stars \#11 and 12 have no membership 
information, and there is a lack of (U-B) values, therefore it cannot make any
 prediction about it.
 Polarimetrically,
star\#11 has a high polarization value to be considered a member, and the orientation of
the polarization vector in the star \#12 is very different from the
 $\bar{\theta}_V$ for the cluster. Star \#13 displays intrinsic polarization,
 and its location in the $Q_V$-$U_V$ plane is clearly of a non-member.

\subsubsection{Background Stars}

On the right side of the same figure, it can be seen three stars (\#1,2 and 4)
which are the most polarized in our sample with 
$P_V=$3.36\%, 3.42\% and 3.35\% respectively.
 On the sky plane (Fig. 1), they are located on the west side of the cluster and
 to the south of a big dust cloud. Their polarization angles 
($\theta_V$=6$^{\circ}$.4, 4$^{\circ}$.8, and 8$^{\circ}$.7, respectively)
 are similar to the mean value of the cluster.
 Star 1 was classified as G8III; therefore, it should have $E_{B-V}=0.87$
mag. Originally, it was considered as a member by Pedreros(1987), but its membership
probability is lower than 50\% (Baumgardt et al. 2000). The other two stars
have not membership information. 
The location of these three stars in the $Q_V$ vs. $U_V$ plot shows
 characteristics of background
 stars, their light being polarized for a dust component between
 them and the cluster.

 
\subsection{Polarization efficiency and distribution of interstellar matter }

The ratio ${P_V}/{E_{B-V}}$ is known as the
 "polarization efficiency" of the interstellar medium and it depends mainly
on the alignment efficiency, the magnetic field strength and the
 depolarization due to radiation traversing more
than one dust cloud with different field directions.

   Fig. 6 displays the relation ($P_V$ vs. $E_{B-V}$) that exists between
the reddening and the polarization produced by the dust grains along the line
 of sight to NGC 6124.
 Assuming normal interstellar material characterized by R=3.2,
the empirical upper limit relation for the polarization efficiency given by
$P_{V}=R {A_V} \sim 9E_{B-V}$ (Serkowski et al. 1975) is depicted by
the solid line in this figure. Indeed, this line represents the maximum
efficiency of the polarization produced by the interstellar dust. Likewise,
the dotted line $P_V=3.5{E_{B-V}}^{0.8}$ represents the most recent estimate of
the average efficiency made by Fosalba et al. (2002), valid for $E_{B-V} < 1.0$
mag. The dashed line $P_V=5E_{B-V}$ was drawn as a
reference and represents the observed normal efficiency
of the polarizing properties of dust given by Serkowski et al. (1975).
In this figure the stars are plotted with our polarimetric membership status.

Excesses $E_{B-V}$ were obtained from the literature or from the relation
between spectral type and colour index following Schmidt-Kaler (1982).
In Fig. 6 the majority of stars lie on the right of the interstellar
maximum efficiency line (${P_V}/{E_{B-V}} \sim 9$, Serkowski et al., 1975), indicating that the observed
 polarization is mostly due to the ISM. Only two stars (\#3,28) lie
on the left of the line; this can be
 associated with errors in the estimate of their excesses.


To derive the mean polarimetric efficiency, we made use of the same 7 stars that we
selected to calculate the mean polarization and angle of NGC 6124 (section
4.2), located around the cross symbol in Fig. 6.
We obtained a mean efficiency ${P_V}/{E_{B-V}}$=3.1$\pm $0.62,
 which is smaller than the
value for the interstellar dust given by the Fosalba's estimate
 (${{P_V}/{E_{B-V}}}\sim 3.66$) and still much smaller
 than the value given
by the Serkowski's estimate (${{P_V}/{E_{B-V}}}\sim 5.$) for the average efficiency..
 This depolarization in the cluster can be the result
of the superposition of several dust clouds on the line of sight 
with different magnetic field directions, confirming the information given in section 4.1. 
Also, it shows scattering in colour excess (approximately $0.60<E_{B-V}<0.89$) and 
in the polarization values ($\Delta P_V\sim$ 1.3\%) between members of the cluster. 
This may be a consequence of the presence of intracluster dust.


Star \#32 on the lower side of the figure is probably a late-type star, whose
excess may be overestimated. This star is marked in the plot using
an arrow that points to smaller excesses.
Stars \#1 and \#33 show low $E_{B-V}$s according to the polarization values. Polarimetrically,
they are non-member stars with erroneous $E_{B-V}$s due to their bad photograficphotometry. 
 Again, as mentioned
in the preceding section, star \#22 shows polarimetric characteristics of a member
of the cluster.

	In all stars without membership data it was not possible to calculate
 their colour excesses. It is the situation
for \#2,7,10,11,12,17,23,24,25,27,31, and 46. In the case of stars \#7, 12, and 24,
they may be considered new probable members of NGC 6124, as their
$P_V$ values are similar to those associated with cluster stars. 
Also, they are located in the region of the $Q_V$ vs. $U_V$ plane where there is
a number of members. The rest of stars may be probably in the background 
(like 2 and 11) or nearby non-members (between the Sun and the cluster) like
 \#10,17,23,25,27,31, and 46.

\subsection{Dust clouds}

Our data show that the orientation of the polarimetric vectors is not parallel to the
 Galactic Plane (Axon et al., 1976), and that there is a considerable degree of depolarization.
These two properties can be explained by the composition of the individual effects of several clouds
along the line of sight to the cluster, where at least one of these layers of dust
polarizes the starlight in an orientation different than the rest, resulting in
depolarization. That unusual component might have been recently perturbed so that their dust
particles might have not had enough time to relax and to orient in the direction of the Galactic Plane.
The most likely candidate to explain this polarimetric component (and its properties) is the Lupus cloud.

NGC 6124 lies behind the Lupus cloud which is a large structure composed of
 several subclouds showing
different modes of star formation. Applying a maximum-likelihood technique to
Hipparcus data, Lombardi et al. (2008)
obtained a distance of d=155$\pm$8 pc, confirming the 
historical estimate of d$\sim$150 pc. The subcloud known
 as Lupus 6 is close to
 the direction of NGC 6124. The cluster is behind a long dark filament that
 crosses Lupus 6 and continues up to Lupus 4.
Feinstein et al. (2003) have shown that, over the region of NGC 6231, the Lupus cloud does no polarize  
the light parallel to the Galactic Plane.
The Lupus cloud also stands out as a strong and compact perturbation of the optical polarization of the Galaxy
in Figs. 1b and 1c of Axon et al. (1976).


Three non-member stars in the region which seem to be polarized by the Lupus cloud, 
(\#16, 20, and 28) give the mean values $\bar{P_V} \sim$ 1.6\% and $\bar{\theta}_V \sim $0$^\circ$.
The combination of this orientation and that of the Galactic Plane ($\theta_V \sim$ 44$^\circ$),
close to 90$^\circ$ in the $Q_V-U_V$ plot, could explain the high depolarization of the data.
Besides, it could explain the spread of the observations along the $Q_V$ axis
 (where $\theta_V \sim$ 0$^\circ$), for all lowly polarized stars. 
As expected, the polarization caused by the Lupus cloud on NGC 6124 is not exactly the
same  as that for NGC 6231, because the cloud changes its optical properties between the two clusters.
In the case of  NGC 6231, the observed polarizations for stars \#102 and 189 (suspected to be
polarized by Lupus), are $P_V$=0.85\% $\pm$ 0.10, $\theta_V$=8$^\circ$.0  $\pm$ 3.3 and $P_V$=0.46\% $\pm$ 0.11,
$\theta_V$=18$^\circ$.4  $\pm$ 6.9 respectively, i.e. somewhat similar in angle, but lower in polarization
percentage, to the mean results for the three non-member stars in the NGC 6124 region.


Fig. 7 shows the trend of  absorption ($\bar{A}_V$) with increasing
distance in the area of NGC 6124. Only stars whose distances were obtained by us
are included, and the dot-dashed line accounts for the presence of three
absorption sheets between the Sun and NGC 6124 (d$\sim$560 pc). The first
dust component is located very near the Sun and affects stars
\#3 and \#32, but does not produce much extinction ($A_V$=0.2). Farther out,
the absorption increases at about 160 pc  (second absorption sheet) as it could be
expected from the presence of the Lupus cloud where stars \#20
 and \#28 are located ($A_V \sim$ 0.5 mag.). Between approximately
200 and 400 pc there are no stars with known distances, but from 400 pc
up to the cluster location Fig. 7 shows that the extinction increases considerably,
perhaps due to the presence of several
dust clouds along the line of sight to the cluster, as mentioned in the preceding sections. 
A step in the absorption value within the cluster itself can be also noticed and it is probably
caused by the presence of an intracluster dust component.


\section{Summary}

 We obtained the linear polarization of a sample of 46 stars in the region of the
open cluster NGC 6124. Through the analysis of our polarimetric data, we
have found evidence for at least three dust layers along the line of sight to the
cluster.

The average polarization has an angle of ${\bar\theta_V}$=8$^{o}$.2 which is not parallel to 
the Galactic Plane, and the mean polarization 
efficiency is ${P_{V}}/{E_{B-V}}$=3.1. This value is lower than the average 
polarization efficiency for the interstellar medium, which  means that the 
cluster is highly depolarized.
The scatter in polarization ($\Delta P_V \sim 1.3 \%$) and in extinction ($\Delta E_{B-V}\sim 0.29$ mag.)
for the members of the cluster are compatible with the presence of intra-cluster dust.

 Our data show that one of these components is associated with the Lupus cloud, and that 
it polarized the light close to $\theta_V$=0$^\circ$. This orientation differs from the Galactic
Plane projection in the region ($\theta_{GP}$=44$^{\circ}$).
We interpret that these two different directions of polarizations tend to depolarize the light
from the cluster stars, because in the 
$Q_V-U_V$ plane it is basically 2$\theta \sim $90$^\circ$.

We have shown that polarization measurements are useful to detect the non-member
stars that contaminate the sample in the NGC 6124 region, and we have been able to detect
 a few stars with intrinsic polarization.

\section*{Acknowledgments}

We wish to acknowledge the technical support at CASLEO during the observing runs. We also 
acknowledge the use of the Torino Photopolarimeter built at Osservatorio Astronomico di Torino
(Italy) and operated under agreement between Complejo Astron\'omico El Leoncito and Osservatorio Astronomico di Torino. 
We thanks Dr. Juan Carlos Muzzio for reading of the manuscript and for his useful comments that helps to improve this paper.  
This research has made use of the WEBDA database, operated at the Institute for Astronomy of the University of Vienna.



\clearpage
\newpage



\begin{table}
\caption {Results for zero polarizations standards}
\begin{tabular}{l|||||c|||||c|||||c|||||c|||||c}
\noalign{\smallskip}
\hline \hline
\noalign{\smallskip}

 &  U &   B & V &   R &  I  \\
    

Date  & $Q_{U} (\%) \pm \epsilon_{Q_U} $  &  $Q_{B} (\%)\pm \epsilon_{Q_B}$  &    $Q_{V}(\%)\pm  \epsilon_{Q_V}$  &  $Q_{R}(\%) \pm  \epsilon_{Q_R}$  &   $Q_{I}(\%)  \pm \epsilon_{Q_I}$   \\
Number of observations &  $U_{U} (\%) \pm \epsilon_{U_U} $  &  $U_{B} (\%)\pm \epsilon_{U_B}$  &    $U_{V}(\%)\pm  \epsilon_{U_V}$  &  $U_{R}(\%) \pm  \epsilon_{U_R}$  &   $U_{I}(\%)  \pm \epsilon_{U_I}$   \\



\noalign{\smallskip}
\hline
\noalign{\smallskip}


29 April-2 May 2003 
& 0.08  $\pm$ 0.13  & 0.10   $\pm$ 0.12 & 0.02  $\pm$ 0.06 & 0.02  $\pm$ 0.08 & 0.03 

 $\pm$ 0.07 \\ 
12& -0.08  $\pm$ 0.2 & -0.06  $\pm$ 0.13 & -0.06  $\pm$ 0.131 & 0.01  $\pm$ 0.16 & -0.07   $\pm$ 0.15 \\
\noalign{\smallskip}
18-20 April 2004 
& -0.12 $\pm$ 0.09 &	0.06 $\pm$ 0.04 &  -0.02 $\pm$ 0.02 & -0.001 $\pm$ 0.02 & -0.011  $\pm$ 0.05 \\

5& 0.02	$\pm$ 0.15 & 0.03 $\pm$  0.05 &	0.05$\pm$ 0.07 &0.002$\pm$ 0.03 &0.006$\pm$ 0.05  \\

\noalign{\smallskip}
21-24 May 2007 
& -0.24 $\pm$ 0.01 & -0.01 $\pm$ 0.01 &-0.09 $\pm$ 0.07 & -0.03$\pm$ 0.05 & -0.04 $\pm$ 0.05 \\ 
6 & -0.190  $\pm$ 0.03 & 	-0.08  $\pm$ 0.02 & 0.03  $\pm$ 0.06 & 	0.006 $\pm$ 0.07&  0.012  $\pm$ 0.14 \\

\noalign{\smallskip}

\noalign{\smallskip}
\hline
\end{tabular}
\end{table}


\newpage


\begin{table}
\caption {Polarimetric Observations of stars in NGC 6124}
\begin{tabular}{lcc}
\noalign{\smallskip}
\hline \hline
\noalign{\smallskip}

Star Id. & $P_{\lambda} \pm \epsilon_{P}$ &  $ \theta_{\lambda} \pm \epsilon_{\theta}$ \\

Filter & \% & $^\circ$ \\

\noalign{\smallskip}
\hline
\noalign{\smallskip}

 Star   1 (rg)&  & \\ 
U &  2.91 $\pm$ 0.69 &   2.4 $\pm$  6.7\\
B &  4.05 $\pm$ 0.31 &   8.4 $\pm$  2.2\\
V &  3.36 $\pm$ 0.23 &   6.4 $\pm$  2.0\\
R &  3.74 $\pm$ 0.21 &   4.9 $\pm$  1.6\\
I &  3.30 $\pm$ 0.28 &   4.0 $\pm$  2.4\\
 Star   2&  & \\ 
U &  2.95 $\pm$ 0.15 &   3.3 $\pm$  1.4\\
B &  3.67 $\pm$ 0.14 &   3.9 $\pm$  1.1\\
V &  3.42 $\pm$ 0.26 &   4.8 $\pm$  2.2\\
R &  3.44 $\pm$ 0.17 &   4.5 $\pm$  1.4\\
I &  3.18 $\pm$ 0.20 &   5.8 $\pm$  1.8\\
 Star   3 (m)&  & \\ 
U &  0.77 $\pm$ 0.17 &  13.7 $\pm$  6.2\\
B &  0.92 $\pm$ 0.09 &  15.0 $\pm$  2.7\\
V &  0.89 $\pm$ 0.08 &  15.1 $\pm$  2.5\\
R &  0.98 $\pm$ 0.08 &  13.1 $\pm$  2.2\\
I &  0.90 $\pm$ 0.10 &  13.9 $\pm$  3.1\\
 Star   4&  & \\
U &  2.54 $\pm$ 0.30 &  15.2 $\pm$  3.4\\
B &  3.46 $\pm$ 0.16 &  11.2 $\pm$  1.3\\
V &  3.35 $\pm$ 0.14 &   8.7 $\pm$  1.2\\
R &  3.56 $\pm$ 0.10 &   9.4 $\pm$  0.8\\
I &  3.13 $\pm$ 0.12 &   8.8 $\pm$  1.1\\
 Star   5 (mP)&  & \\ 
U &  2.25 $\pm$ 0.22 &  13.3 $\pm$  2.8\\
B &  2.68 $\pm$ 0.17 &  11.2 $\pm$  1.8\\
V &  2.77 $\pm$ 0.15 &  12.2 $\pm$  1.5\\
R &  2.70 $\pm$ 0.12 &  11.4 $\pm$  1.3\\
I &  2.62 $\pm$ 0.15 &  10.4 $\pm$  1.6\\
 Star   6 (nP)&  & \\
U &  0.46 $\pm$ 0.33 &  56.6 $\pm$ 17.7\\
B &  0.23 $\pm$ 0.27 &  83.4 $\pm$ 24.7\\
V &  0.11 $\pm$ 0.34 &  82.6 $\pm$ 36.0\\
R &  0.08 $\pm$ 0.17 &  92.1 $\pm$ 32.0\\
I &  0.14 $\pm$ 0.20 &  66.2 $\pm$ 27.7\\
 Star   7&  & \\ 
U &  2.05 $\pm$ 0.28 &   8.5 $\pm$  3.8\\
B &  2.33 $\pm$ 0.25 &  10.4 $\pm$  3.0\\
V &  2.09 $\pm$ 0.32 &  13.0 $\pm$  4.3\\
R &  2.03 $\pm$ 0.22 &  12.7 $\pm$  3.0\\
I &  1.73 $\pm$ 0.25 &   7.8 $\pm$  4.1\\
 Star   8 (mP)&  & \\ 
U &  1.67 $\pm$ 0.19 &  13.0 $\pm$  3.3\\
B &  2.09 $\pm$ 0.13 &   8.6 $\pm$  1.7\\
V &  1.95 $\pm$ 0.12 &   9.6 $\pm$  1.7\\
R &  2.02 $\pm$ 0.10 &   9.4 $\pm$  1.5\\
I &  1.75 $\pm$ 0.14 &   8.3 $\pm$  2.3\\
 Star   9 (n)&  & \\
U &  0.65 $\pm$ 0.34 &  57.9 $\pm$ 13.8\\
B &  0.19 $\pm$ 0.18 &  42.5 $\pm$ 22.4\\
V &  0.23 $\pm$ 0.13 &  99.6 $\pm$ 14.5\\
R &  0.05 $\pm$ 0.12 & 130.4 $\pm$ 33.0\\
I &  0.09 $\pm$ 0.17 & 160.1 $\pm$ 31.4\\

\noalign{\smallskip}
\hline
\end{tabular}
\end{table}

\addtocounter{table}{-1}%
\begin{table}
\caption{continued}
\begin{tabular}{lcc}

\noalign{\smallskip}
\hline\hline
\noalign{\smallskip}

Star Id. & $P_{\lambda} \pm \epsilon_{P}$ &  $ \theta_{\lambda} \pm \epsilon_{\theta}$ \\

Filter & \% & $^\circ$ \\

\noalign{\smallskip}
\hline
\noalign{\smallskip}

 Star   10&  & \\
U &  1.67 $\pm$ 0.36 &   0.1 $\pm$  6.1\\
B &  1.87 $\pm$ 0.32 &   5.3 $\pm$  4.8\\
V &  1.76 $\pm$ 0.34 &   8.2 $\pm$  5.5\\
R &  1.71 $\pm$ 0.21 &  11.3 $\pm$  3.5\\
I &  1.62 $\pm$ 0.29 &  13.5 $\pm$  5.1\\
 Star   11&  & \\ 
U &  2.60 $\pm$ 0.22 & 179.6 $\pm$  2.4\\
B &  2.88 $\pm$ 0.10 &   0.6 $\pm$  1.0\\
V &  3.03 $\pm$ 0.15 & 178.9 $\pm$  1.4\\
R &  3.04 $\pm$ 0.11 & 178.8 $\pm$  1.0\\
I &  2.72 $\pm$ 0.17 & 177.9 $\pm$  1.8\\
 Star   12&  & \\
U &  2.06 $\pm$ 0.56 &   5.9 $\pm$  7.6\\
B &  2.06 $\pm$ 0.17 & 177.3 $\pm$  2.3\\
V &  2.44 $\pm$ 0.27 & 173.6 $\pm$  3.2\\
R &  2.48 $\pm$ 0.23 & 174.6 $\pm$  2.7\\
I &  2.05 $\pm$ 0.27 & 174.6 $\pm$  3.8\\
 Star   13 (nP)&  & \\ 
U &  1.53 $\pm$ 0.25 &   4.1 $\pm$  4.6\\
B &  1.84 $\pm$ 0.19 &   1.4 $\pm$  3.0\\
V &  1.76 $\pm$ 0.22 & 170.5 $\pm$  3.6\\
R &  1.90 $\pm$ 0.19 & 172.3 $\pm$  2.9\\
I &  1.65 $\pm$ 0.22 & 168.7 $\pm$  3.7\\
 Star   14 (rg)&  & \\ 
U &  3.41 $\pm$ 0.87 &  12.0 $\pm$  7.1\\
B &  2.44 $\pm$ 0.29 &   7.6 $\pm$  3.4\\
V &  2.52 $\pm$ 0.26 &   5.8 $\pm$  3.0\\
R &  2.64 $\pm$ 0.22 &   7.4 $\pm$  2.4\\
I &  2.18 $\pm$ 0.28 &   6.5 $\pm$  3.6\\
 Star   15 (mP)&  & \\ 
U &  2.29 $\pm$ 0.28 & 178.6 $\pm$  3.5\\
B &  2.44 $\pm$ 0.16 &   3.0 $\pm$  1.8\\
V &  2.53 $\pm$ 0.18 &   1.6 $\pm$  2.0\\
R &  2.68 $\pm$ 0.13 &   2.2 $\pm$  1.4\\
I &  2.25 $\pm$ 0.16 &   0.3 $\pm$  2.1\\
 Star   16 (mP)&  & \\
U &  1.31 $\pm$ 0.30 & 176.6 $\pm$  6.5\\
B &  1.58 $\pm$ 0.09 & 177.6 $\pm$  1.6\\
V &  1.54 $\pm$ 0.11 & 176.0 $\pm$  2.1\\
R &  1.60 $\pm$ 0.09 & 175.1 $\pm$  1.6\\
I &  1.41 $\pm$ 0.13 & 173.9 $\pm$  2.6\\
 Star   17&  & \\
U &  1.94 $\pm$ 0.51 &   6.6 $\pm$  7.3\\
B &  1.75 $\pm$ 0.25 &   1.9 $\pm$  4.0\\
V &  1.59 $\pm$ 0.15 &   5.9 $\pm$  2.7\\
R &  1.56 $\pm$ 0.10 &   2.4 $\pm$  1.9\\
I &  1.33 $\pm$ 0.18 &   2.1 $\pm$  3.9\\
 Star   18 (mP)&  & \\ 
U &  1.64 $\pm$ 0.19 &   7.6 $\pm$  3.3\\
B &  1.87 $\pm$ 0.15 &   7.7 $\pm$  2.3\\
V &  1.89 $\pm$ 0.18 &   7.5 $\pm$  2.7\\
R &  1.90 $\pm$ 0.15 &   5.9 $\pm$  2.2\\
I &  1.36 $\pm$ 0.20 &   6.1 $\pm$  4.3\\

\noalign{\smallskip}
\hline
\end{tabular}
\end{table}

\addtocounter{table}{-1}%
\begin{table}
\caption{continued}
\begin{tabular}{lcc}

\noalign{\smallskip}
\hline\hline
\noalign{\smallskip}

Star Id. & $P_{\lambda} \pm \epsilon_{P}$ &  $ \theta_{\lambda} \pm \epsilon_{\theta}$ \\

Filter & \% & $^\circ$ \\

\noalign{\smallskip}
\hline
\noalign{\smallskip}

 Star   19&  & \\
U &  1.29 $\pm$ 0.31 & 178.3 $\pm$  6.6\\
B &  2.32 $\pm$ 0.10 & 177.5 $\pm$  1.3\\
V &  2.30 $\pm$ 0.12 & 175.4 $\pm$  1.5\\
R &  2.43 $\pm$ 0.09 & 175.4 $\pm$  1.0\\
I &  2.11 $\pm$ 0.11 & 176.0 $\pm$  1.5\\
 Star   20 (nP)&  & \\ 
U &  1.36 $\pm$ 0.23 &   7.1 $\pm$  4.8\\
B &  1.67 $\pm$ 0.13 &   4.2 $\pm$  2.3\\
V &  1.57 $\pm$ 0.14 &   2.7 $\pm$  2.5\\
R &  1.61 $\pm$ 0.12 &   2.6 $\pm$  2.2\\
I &  1.48 $\pm$ 0.16 &   0.3 $\pm$  3.2\\
 Star   21 (mP)&  & \\
U &  1.94 $\pm$ 0.35 &   5.1 $\pm$  5.1\\
B &  2.33 $\pm$ 0.10 &   5.0 $\pm$  1.3\\
V &  2.27 $\pm$ 0.10 &   3.2 $\pm$  1.2\\
R &  2.40 $\pm$ 0.06 &   4.6 $\pm$  0.8\\
I &  2.24 $\pm$ 0.09 &   5.6 $\pm$  1.2\\
 Star   22 (nP)&  & \\
U &  2.06 $\pm$ 0.47 &   7.4 $\pm$  6.5\\
B &  2.36 $\pm$ 0.16 &   3.2 $\pm$  1.9\\
V &  2.41 $\pm$ 0.15 &   5.5 $\pm$  1.8\\
R &  2.46 $\pm$ 0.13 &   5.0 $\pm$  1.5\\
I &  2.32 $\pm$ 0.15 &   5.2 $\pm$  1.9\\
 Star   23 (P)&  & \\ 
U &  1.53 $\pm$ 0.20 &   4.9 $\pm$  3.8\\
B &  1.65 $\pm$ 0.12 &   7.4 $\pm$  2.1\\
V &  1.82 $\pm$ 0.13 &   5.3 $\pm$  2.1\\
R &  1.76 $\pm$ 0.09 &   3.8 $\pm$  1.5\\
I &  1.60 $\pm$ 0.13 &   1.8 $\pm$  2.4\\
 Star   24 (mP)&  & \\ 
U &  1.63 $\pm$ 0.21 &   0.0 $\pm$  3.6\\
B &  1.95 $\pm$ 0.16 &   3.5 $\pm$  2.4\\
V &  2.05 $\pm$ 0.12 &   4.0 $\pm$  1.6\\
R &  2.07 $\pm$ 0.13 &   2.4 $\pm$  1.8\\
I &  1.75 $\pm$ 0.18 &   1.8 $\pm$  3.0\\
 Star   25&  & \\ 
U &  1.48 $\pm$ 0.16 &  10.8 $\pm$  3.0\\
B &  1.71 $\pm$ 0.07 &  12.4 $\pm$  1.2\\
V &  1.65 $\pm$ 0.08 &  12.2 $\pm$  1.4\\
R &  1.72 $\pm$ 0.06 &  10.7 $\pm$  0.9\\
I &  1.43 $\pm$ 0.10 &  10.6 $\pm$  2.0\\
 Star   26 (nP)&  & \\ 
U &  2.08 $\pm$ 0.33 & 179.8 $\pm$  4.5\\
B &  2.46 $\pm$ 0.30 &   7.0 $\pm$  3.5\\
V &  2.44 $\pm$ 0.40 &  11.9 $\pm$  4.6\\
R &  2.30 $\pm$ 0.34 &  11.1 $\pm$  4.2\\
I &  2.09 $\pm$ 0.33 &  12.1 $\pm$  4.5\\
 Star   27&  & \\ 
U &  0.88 $\pm$ 0.19 &  19.8 $\pm$  5.9\\
B &  1.05 $\pm$ 0.10 &   9.7 $\pm$  2.7\\
V &  1.19 $\pm$ 0.12 &   6.3 $\pm$  3.0\\
R &  1.25 $\pm$ 0.16 &   5.6 $\pm$  3.7\\
I &  1.08 $\pm$ 0.17 &   3.4 $\pm$  4.5\\

\noalign{\smallskip}
\hline
\end{tabular}
\end{table}

\addtocounter{table}{-1}%
\begin{table}
\caption{continued}
\begin{tabular}{lcc}

\noalign{\smallskip}
\hline\hline
\noalign{\smallskip}

Star Id. & $P_{\lambda} \pm \epsilon_{P}$ &  $ \theta_{\lambda} \pm \epsilon_{\theta}$ \\

Filter & \% & $^\circ$ \\

\noalign{\smallskip}
\hline
\noalign{\smallskip}

 Star   28 (nP)&  & \\
U &  1.91 $\pm$ 0.37 &   3.9 $\pm$  5.4\\
B &  1.88 $\pm$ 0.15 &   3.2 $\pm$  2.3\\
V &  1.84 $\pm$ 0.14 &   0.8 $\pm$  2.3\\
R &  1.85 $\pm$ 0.09 &   1.8 $\pm$  1.4\\
I &  1.68 $\pm$ 0.12 &   4.4 $\pm$  2.1\\
 Star   29 (rgP)&  & \\ 
U &  1.55 $\pm$ 1.27 &  34.2 $\pm$ 19.6\\
B &  2.07 $\pm$ 0.30 &   2.5 $\pm$  4.1\\
V &  2.22 $\pm$ 0.17 &   2.9 $\pm$  2.2\\
R &  2.24 $\pm$ 0.20 &   2.4 $\pm$  2.5\\
I &  1.89 $\pm$ 0.21 &   1.5 $\pm$  3.2\\
 Star   30 (m)&  & \\ 
U &  2.09 $\pm$ 0.34 &   5.3 $\pm$  4.6\\
B &  2.49 $\pm$ 0.20 &   9.5 $\pm$  2.3\\
V &  2.46 $\pm$ 0.15 &   6.4 $\pm$  1.7\\
R &  2.45 $\pm$ 0.13 &   9.1 $\pm$  1.5\\
I &  1.98 $\pm$ 0.26 &  10.7 $\pm$  3.7\\
 Star   31&  & \\
U &  0.54 $\pm$ 0.34 &  32.2 $\pm$ 16.1\\
B &  0.03 $\pm$ 0.12 &  25.3 $\pm$ 39.2\\
V &  0.07 $\pm$ 0.13 &  77.6 $\pm$ 30.4\\
R &  0.06 $\pm$ 0.14 &  70.1 $\pm$ 33.5\\
I &  0.16 $\pm$ 0.15 &  99.3 $\pm$ 21.1\\
 Star   32 (mP)&  & \\ 
U &  0.94 $\pm$ 0.36 &  16.5 $\pm$ 10.6\\
B &  0.97 $\pm$ 0.22 &  13.5 $\pm$  6.2\\
V &  1.02 $\pm$ 0.14 &   6.9 $\pm$  3.9\\
R &  1.04 $\pm$ 0.10 &  11.0 $\pm$  2.6\\
I &  0.69 $\pm$ 0.20 &   2.7 $\pm$  8.2\\
 Star   33 (rg)&  & \\ 
U &  2.54 $\pm$ 0.65 &   3.6 $\pm$  7.2\\
B &  2.96 $\pm$ 0.18 &  16.7 $\pm$  1.7\\
V &  2.78 $\pm$ 0.11 &  15.8 $\pm$  1.1\\
R &  2.90 $\pm$ 0.09 &  16.6 $\pm$  0.9\\
I &  2.48 $\pm$ 0.09 &  15.4 $\pm$  1.0\\
 Star   34 (n)&  & \\ 
U &  1.64 $\pm$ 0.25 &  22.6 $\pm$  4.4\\
B &  2.66 $\pm$ 0.13 &  14.7 $\pm$  1.4\\
V &  2.46 $\pm$ 0.13 &  17.2 $\pm$  1.5\\
R &  2.65 $\pm$ 0.12 &  16.1 $\pm$  1.3\\
I &  2.24 $\pm$ 0.16 &  15.3 $\pm$  2.1\\
 Star   35 (rg)&  & \\ 
U &  1.49 $\pm$ 0.49 &   1.6 $\pm$  9.0\\
B &  2.48 $\pm$ 0.11 &   2.8 $\pm$  1.3\\
V &  2.39 $\pm$ 0.11 &   4.9 $\pm$  1.3\\
R &  2.49 $\pm$ 0.08 &   4.4 $\pm$  0.9\\
I &  2.18 $\pm$ 0.10 &   4.0 $\pm$  1.3\\
 Star   36 (rg)&  & \\ 
U &  4.52 $\pm$ 1.25 &  16.0 $\pm$  7.7\\
B &  2.38 $\pm$ 0.25 &  20.7 $\pm$  3.0\\
V &  2.17 $\pm$ 0.20 &  28.2 $\pm$  2.6\\
R &  2.33 $\pm$ 0.22 &  27.9 $\pm$  2.7\\
I &  2.12 $\pm$ 0.23 &  31.7 $\pm$  3.1\\

\noalign{\smallskip}
\hline
\end{tabular}
\end{table}

\addtocounter{table}{-1}%
\begin{table}
\caption{continued}
\begin{tabular}{lcc}

\noalign{\smallskip}
\hline\hline
\noalign{\smallskip}

Star Id. & $P_{\lambda} \pm \epsilon_{P}$ &  $ \theta_{\lambda} \pm \epsilon_{\theta}$ \\

Filter & \% & $^\circ$ \\

\noalign{\smallskip}
\hline
\noalign{\smallskip}

 Star   37 (mP)&  & \\ 
U &  1.84 $\pm$ 0.37 &  13.0 $\pm$  5.6\\
B &  2.27 $\pm$ 0.21 &  11.9 $\pm$  2.6\\
V &  2.40 $\pm$ 0.13 &  10.1 $\pm$  1.6\\
R &  2.39 $\pm$ 0.11 &   8.0 $\pm$  1.3\\
I &  2.25 $\pm$ 0.17 &   6.5 $\pm$  2.2\\
 Star   38 (mP)&  & \\ 
U &  1.21 $\pm$ 0.21 &   3.5 $\pm$  4.9\\
B &  1.51 $\pm$ 0.12 &   9.7 $\pm$  2.3\\
V &  1.71 $\pm$ 0.09 &   9.0 $\pm$  1.6\\
R &  1.64 $\pm$ 0.08 &   9.8 $\pm$  1.4\\
I &  1.58 $\pm$ 0.14 &   7.2 $\pm$  2.4\\
 Star   39 (nP)&  & \\ 
U &  1.74 $\pm$ 0.19 &  18.6 $\pm$  3.2\\
B &  2.20 $\pm$ 0.10 &  19.7 $\pm$  1.3\\
V &  2.11 $\pm$ 0.09 &  17.2 $\pm$  1.2\\
R &  2.20 $\pm$ 0.07 &  17.2 $\pm$  0.9\\
I &  2.06 $\pm$ 0.09 &  15.2 $\pm$  1.3\\
 Star   40 (mP)&  & \\
U &  2.66 $\pm$ 0.44 &  15.7 $\pm$  4.6\\
B &  1.91 $\pm$ 0.17 &  11.5 $\pm$  2.6\\
V &  2.07 $\pm$ 0.11 &  11.7 $\pm$  1.5\\
R &  2.10 $\pm$ 0.07 &  11.0 $\pm$  0.9\\
I &  1.84 $\pm$ 0.15 &  11.3 $\pm$  2.4\\
 Star   41 (rg)&  & \\ 
U &  2.31 $\pm$ 1.15 &  22.7 $\pm$ 13.2\\
B &  1.85 $\pm$ 0.27 &   5.6 $\pm$  4.2\\
V &  1.58 $\pm$ 0.23 &  10.2 $\pm$  4.1\\
R &  1.69 $\pm$ 0.26 &  11.5 $\pm$  4.4\\
I &  1.58 $\pm$ 0.28 &  11.5 $\pm$  5.1\\
 Star   42 (mP)&  & \\ 
U &  2.21 $\pm$ 0.22 &  17.2 $\pm$  2.9\\
B &  2.45 $\pm$ 0.18 &  18.7 $\pm$  2.1\\
V &  2.13 $\pm$ 0.08 &  17.6 $\pm$  1.0\\
R &  2.27 $\pm$ 0.11 &  17.2 $\pm$  1.4\\
I &  2.13 $\pm$ 0.11 &  16.4 $\pm$  1.5\\
 Star   43 (m)&  & \\ 
U &  2.26 $\pm$ 0.21 &  14.7 $\pm$  2.6\\
B &  2.82 $\pm$ 0.18 &  17.7 $\pm$  1.8\\
V &  2.97 $\pm$ 0.12 &  18.1 $\pm$  1.2\\
R &  3.04 $\pm$ 0.11 &  17.9 $\pm$  1.0\\
I &  2.77 $\pm$ 0.16 &  19.1 $\pm$  1.7\\
 Star   44 (n)&  & \\ 
U &  1.32 $\pm$ 0.16 &  14.7 $\pm$  3.5\\
B &  1.80 $\pm$ 0.12 &  12.0 $\pm$  1.8\\
V &  1.71 $\pm$ 0.23 &  15.3 $\pm$  3.8\\
R &  1.82 $\pm$ 0.20 &  15.3 $\pm$  3.1\\
I &  1.63 $\pm$ 0.15 &  17.0 $\pm$  2.7\\
 Star   45 (n)&  & \\ 
U &  1.12 $\pm$ 0.31 &  20.0 $\pm$  7.8\\
B &  1.56 $\pm$ 0.17 &  19.7 $\pm$  3.0\\
V &  1.62 $\pm$ 0.14 &  18.6 $\pm$  2.5\\
R &  1.67 $\pm$ 0.12 &  20.2 $\pm$  2.1\\
I &  1.38 $\pm$ 0.23 &  16.0 $\pm$  4.8\\

\noalign{\smallskip}
\hline
\end{tabular}
\end{table}

\addtocounter{table}{-1}%
\begin{table}
\caption{continued}
\begin{tabular}{lcc}

\noalign{\smallskip}
\hline\hline
\noalign{\smallskip}

Star Id. & $P_{\lambda} \pm \epsilon_{P}$ &  $ \theta_{\lambda} \pm \epsilon_{\theta}$ \\

Filter & \% & $^\circ$ \\

\noalign{\smallskip}
\hline
\noalign{\smallskip}

 Star   46&  & \\ 
U &  1.50 $\pm$ 0.18 &  34.0 $\pm$  3.5\\
B &  1.73 $\pm$ 0.12 &  24.4 $\pm$  1.9\\
V &  1.78 $\pm$ 0.17 &  20.0 $\pm$  2.7\\
R &  1.86 $\pm$ 0.13 &  20.7 $\pm$  2.0\\
I &  1.70 $\pm$ 0.16 &  17.0 $\pm$  2.7\\

\noalign{\smallskip}
\hline
\end{tabular}

\noindent {m: member (determination from the literature, see text)}

\noindent {n: non-member}

\noindent {rg: red giant}

\noindent {P: observed by Pedreros (1987)}

\end{table}

~            



\begin{table}

\caption{Parameters of the Serkowski's fit to the linear polarization data for stars in NGC 6124 }
\begin{tabular}{cccccr}
\noalign{\smallskip}
\hline \hline
\noalign{\smallskip}

Stellar & $P_{max} \pm \epsilon_{P}$ &  $\sigma_{1}^a$ &$\lambda_{max} \pm \epsilon_{\lambda_{max}}$ & $\bar{\epsilon}$   \\
Identification & $\%$ & & m$\mu$ & \\
  
\noalign{\smallskip}
\hline
\noalign{\smallskip}

1  &  3.753 $\pm$   0.182&  1.392&  0.567$\pm$   0.063&   0.85\\
2  &  3.636 $\pm$   0.090&  1.061&  0.559$\pm$   0.024&   0.38\\
3  &  0.974 $\pm$   0.024&  0.543&  0.614$\pm$   0.033&   0.23\\
4  &  3.538 $\pm$   0.088&  1.266&  0.595$\pm$   0.033&   0.93\\
5  &  2.826 $\pm$   0.035&  0.494&  0.592$\pm$   0.015&   0.34\\
7  &  2.252 $\pm$   0.047&  0.354&  0.483$\pm$   0.016&   1.06\\
8  &  2.062 $\pm$   0.051&  0.849&  0.552$\pm$   0.029&   0.55\\
10  &  1.867 $\pm$   0.040&  0.266&  0.526$\pm$   0.021&   3.51\\
11  &  3.093 $\pm$   0.020&  0.312&  0.576$\pm$   0.008&   0.58\\
12  &  2.366 $\pm$   0.082&  0.641&  0.617$\pm$   0.042&   1.49\\
13  &  1.894 $\pm$   0.043&  0.428&  0.576$\pm$   0.025&   9.06\\
14  &  2.646 $\pm$   0.127&  0.901&  0.539$\pm$   0.054&   0.40\\
15  &  2.644 $\pm$   0.059&  0.745&  0.572$\pm$   0.026&   0.62\\
16  &  1.633 $\pm$   0.029&  0.541&  0.562$\pm$   0.022&   0.57\\
17  &  1.714 $\pm$   0.075&  0.620&  0.488$\pm$   0.036&   0.70\\
18  &  1.895 $\pm$   0.070&  0.781&  0.512$\pm$   0.037&   0.22\\
19  &  2.400 $\pm$   0.101&  1.489&  0.605$\pm$   0.057&   0.33\\
20  &  1.671 $\pm$   0.040&  0.587&  0.560$\pm$   0.029&   0.85\\
21  &  2.435 $\pm$   0.043&  0.943&  0.599$\pm$   0.025&   0.30\\
22  &  2.518 $\pm$   0.027&  0.359&  0.599$\pm$   0.015&   0.29\\
23  &  1.809 $\pm$   0.017&  0.274&  0.575$\pm$   0.011&   1.28\\
24  &  2.065 $\pm$   0.032&  0.430&  0.581$\pm$   0.020&   0.53\\
25  &  1.749 $\pm$   0.038&  0.970&  0.546$\pm$   0.026&   0.57\\
26  &  2.465 $\pm$   0.031&  0.192&  0.539$\pm$   0.012&   4.84\\
27  &  1.206 $\pm$   0.024&  0.283&  0.642$\pm$   0.021&   3.85\\
28  &  1.942 $\pm$   0.044&  0.588&  0.546$\pm$   0.025&   0.58\\
29  &  2.232 $\pm$   0.048&  0.393&  0.564$\pm$   0.028&   0.83\\
30  &  2.514 $\pm$   0.053&  0.558&  0.538$\pm$   0.026&   0.81\\
32  &  1.050 $\pm$   0.099&  0.799&  0.516$\pm$   0.099&   1.41\\
33  &  2.930 $\pm$   0.084&  1.181&  0.551$\pm$   0.031&   0.44\\
34  &  2.613 $\pm$   0.130&  1.715&  0.596$\pm$   0.065&   1.12\\
35  &  2.535 $\pm$   0.065&  1.128&  0.559$\pm$   0.031&   0.42\\
36  &  2.355 $\pm$   0.094&  0.796&  0.576$\pm$   0.054&   5.06\\
37  &  2.441 $\pm$   0.015&  0.185&  0.606$\pm$   0.009&   0.90\\
38  &  1.688 $\pm$   0.025&  0.466&  0.623$\pm$   0.022&   0.66\\
39  &  2.247 $\pm$   0.046&  1.081&  0.593$\pm$   0.028&   1.18\\
40  &  2.145 $\pm$   0.078&  1.310&  0.567$\pm$   0.052&   0.23\\
41  &  1.754 $\pm$   0.098&  0.696&  0.553$\pm$   0.071&   1.37\\
42  &  2.301 $\pm$   0.096&  1.810&  0.576$\pm$   0.057&   0.23\\
43  &  3.045 $\pm$   0.030&  0.402&  0.620$\pm$   0.013&   0.49\\
44  &  1.829 $\pm$   0.074&  0.831&  0.602$\pm$   0.040&   1.34\\
45  &  1.648 $\pm$   0.048&  0.532&  0.606$\pm$   0.041&   0.34\\
46  &  1.871 $\pm$   0.018&  0.254&  0.597$\pm$   0.011&   7.96\\
\noalign{\smallskip}
\hline

\end{tabular}
\begin{list}{}{}
\item a) ~$ {\sigma_{1}}^{2} = \sum {(r_{\lambda}/{\epsilon_
{p_\lambda}})^2}/(m-2)$ ; where 
\item {~~}~~~  $m$ is the number of colors and
\item {~~}~~~  $r_{\lambda} = P_{\lambda} - P_{\rm
max}~\exp(-K~{ln}^2({\lambda_{\rm max}/{\lambda}))}$

\end{list}

\end{table}
                                                            

\clearpage
\newpage

\begin{figure*}
        \centering
        \includegraphics[width=15cm,angle=-90]{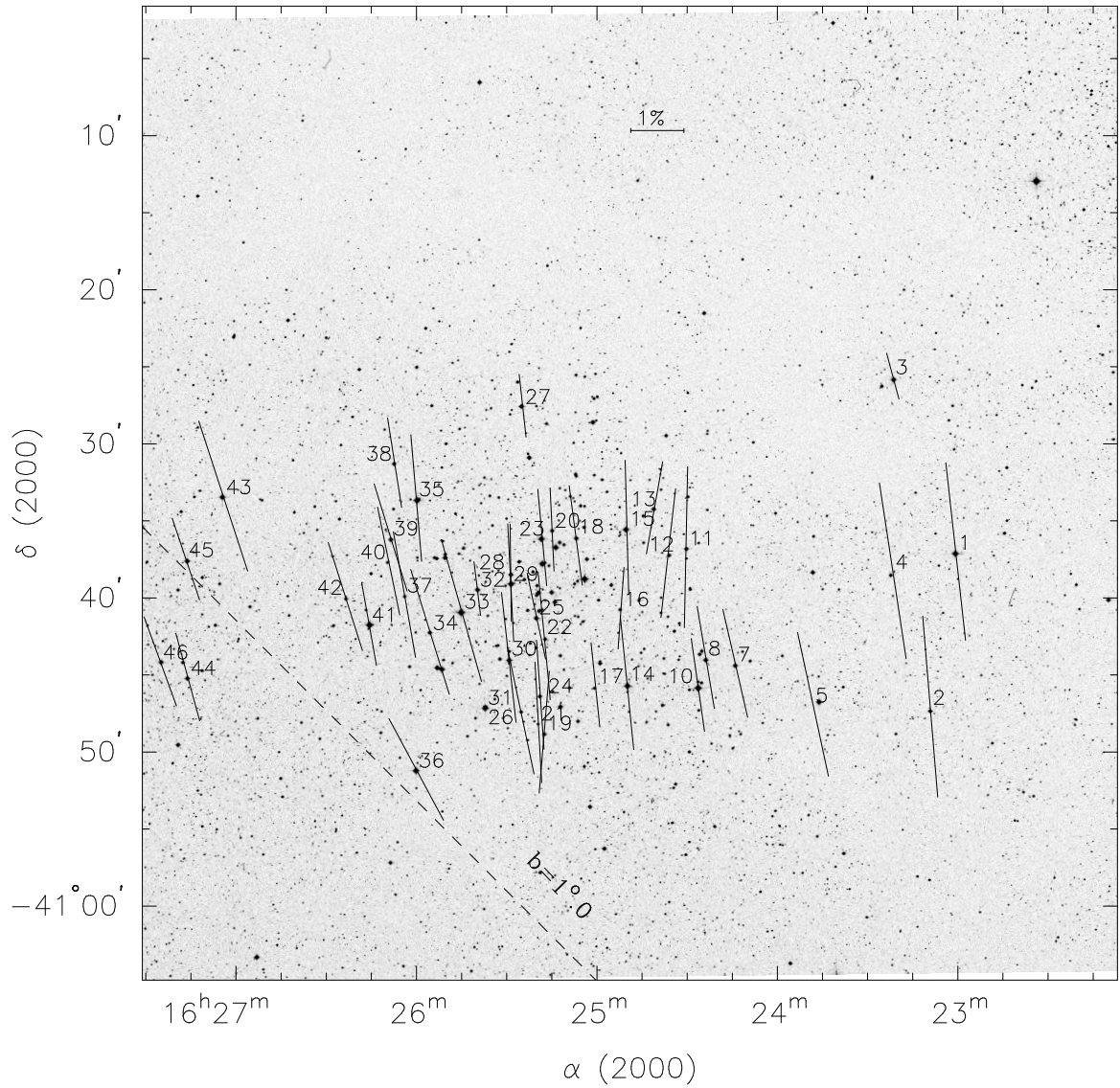}
        \caption{Projection on the sky of the polarization vectors (Johnson V filter) of the stars observed in the region of NGC 6124. The dashed line is the galactic parallel $b=1^{\circ}$0.. This plot shows only the observed stars close to the central core.}
\end{figure*}

\begin{figure*}
        \centering
        \includegraphics[width=11cm]{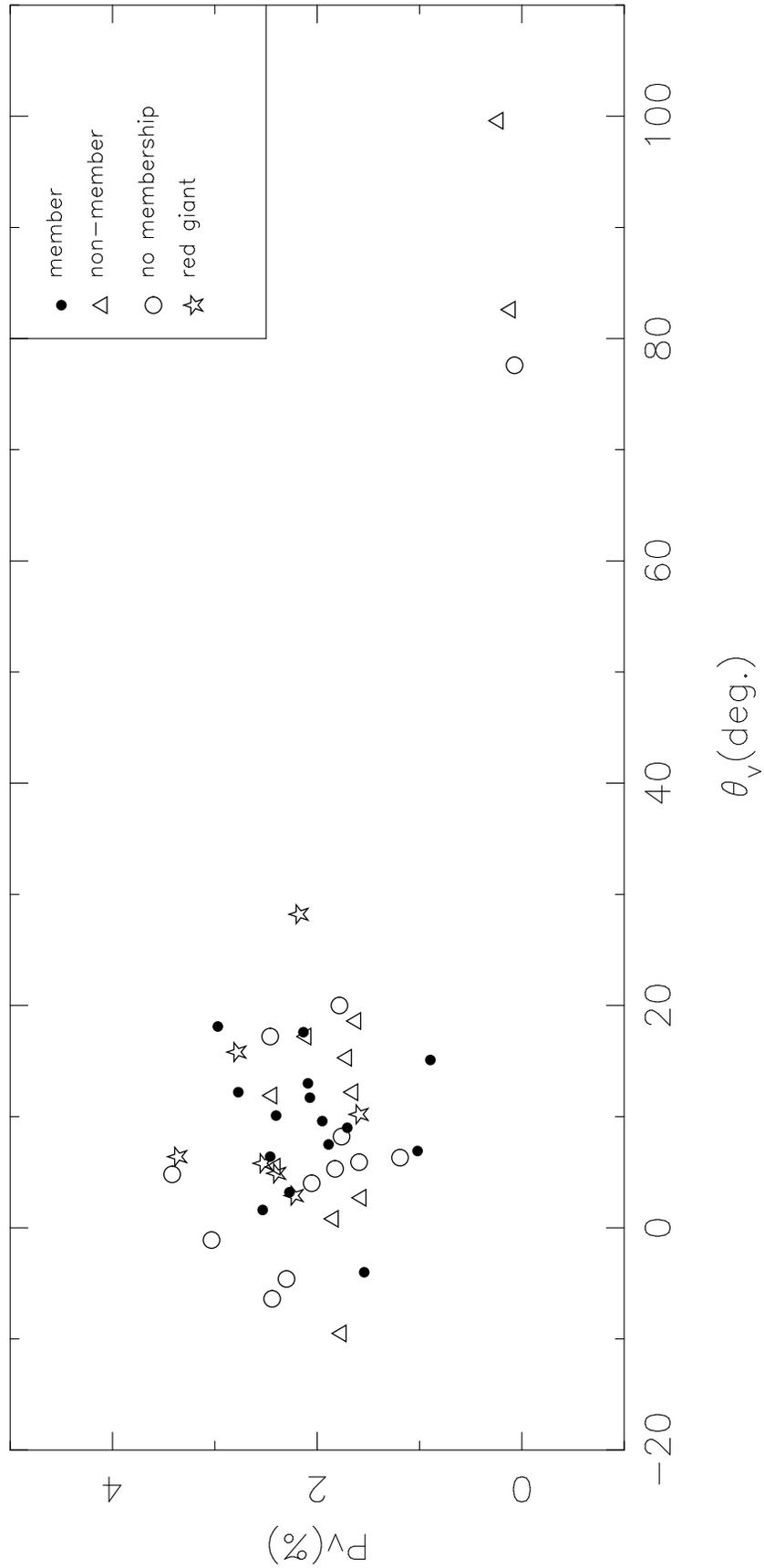}
        \caption{V-band polarization percentage of the stellar flux
 $P_{V}(\%)$ vs. the polarization angle $\theta_{V}$ for each star. Symbols indicate the membership classification from the literature. No membership means that there are not previous membership studies.}
\end{figure*}

\begin{figure*}
        \centering
        \includegraphics[width=13cm,angle=-90]{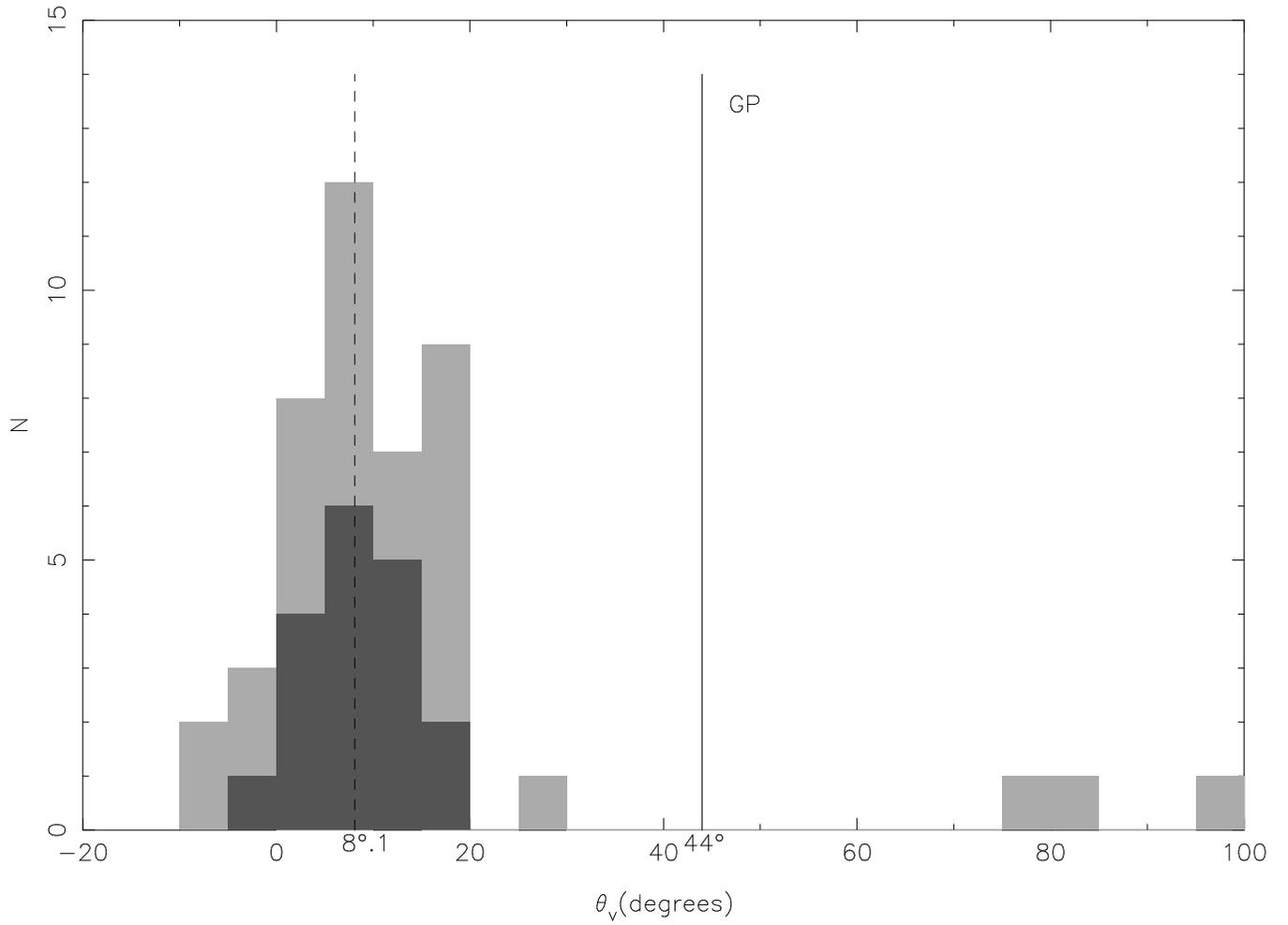}
        \caption{Polarization angle distribution. Grey bars are the histrogram for all the stars observed in the region, while the dark ones are for the member stars. The solid line, labeled as GP, is the projection of the Galactic Plane in the region and the dashed line is the average angle of the polarization vector for the cluster.}
\end{figure*}

\begin{figure*}
        \centering
        \includegraphics[width=13cm]{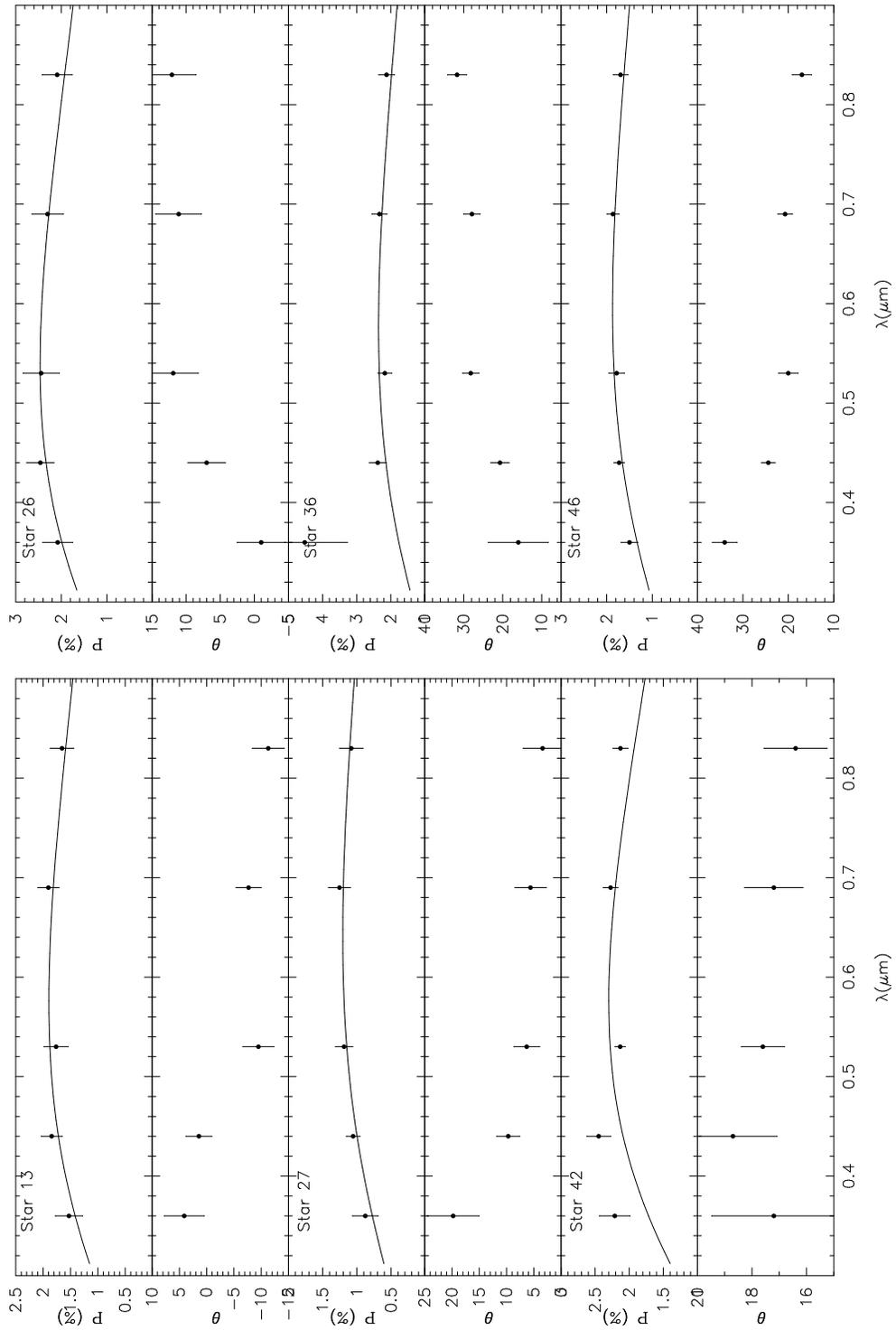}
        \caption{Polarization and position angle dependence on wavelength for stars with features of intrinsic
 polarization}
\end{figure*}

\begin{figure*}
        \centering
        \includegraphics[width=15cm,angle=-90]{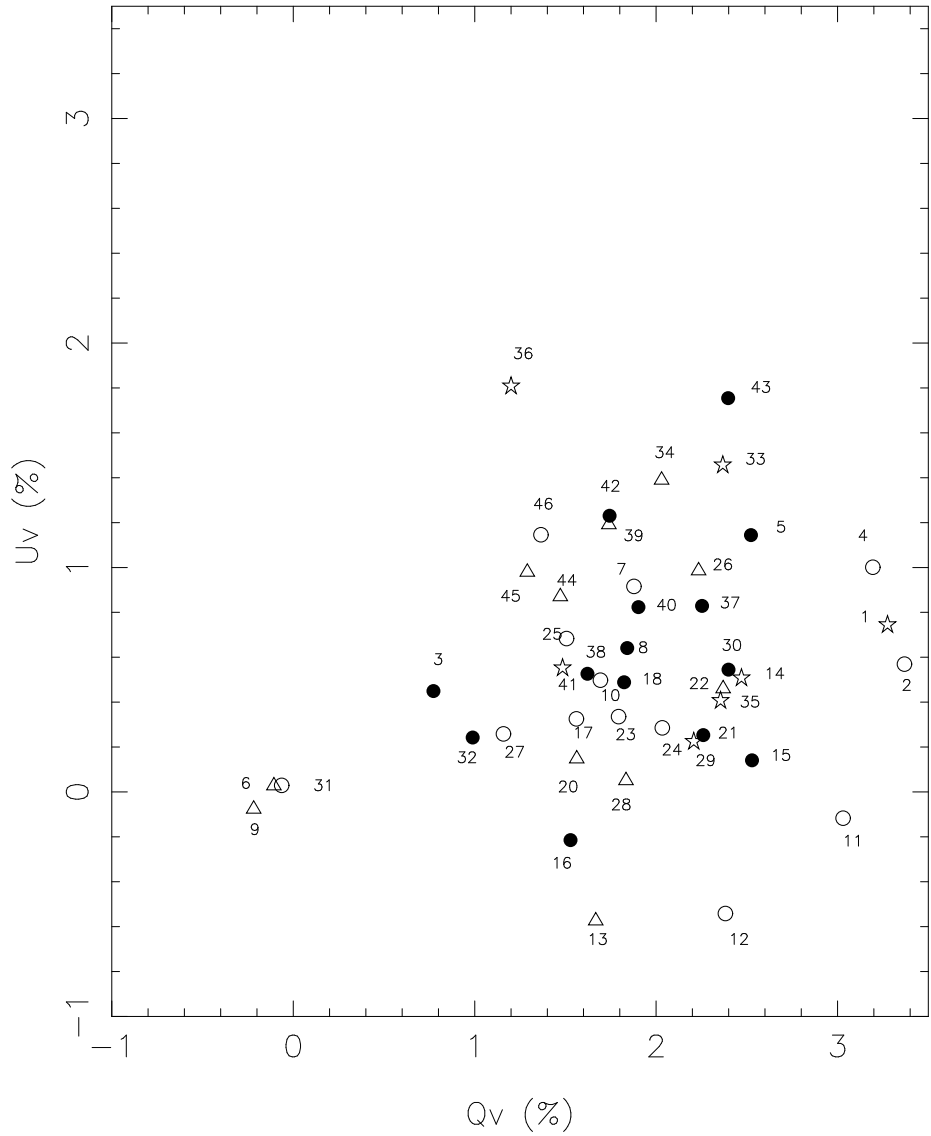}
\caption{ $U_V$ vs. $Q_V$ plot for the stars in the region of NGC 6124. Symbols are as in Fig. 2.}
\label{}
\end{figure*}

\begin{figure*}
        \centering
        \includegraphics[width=17cm,angle=-90]{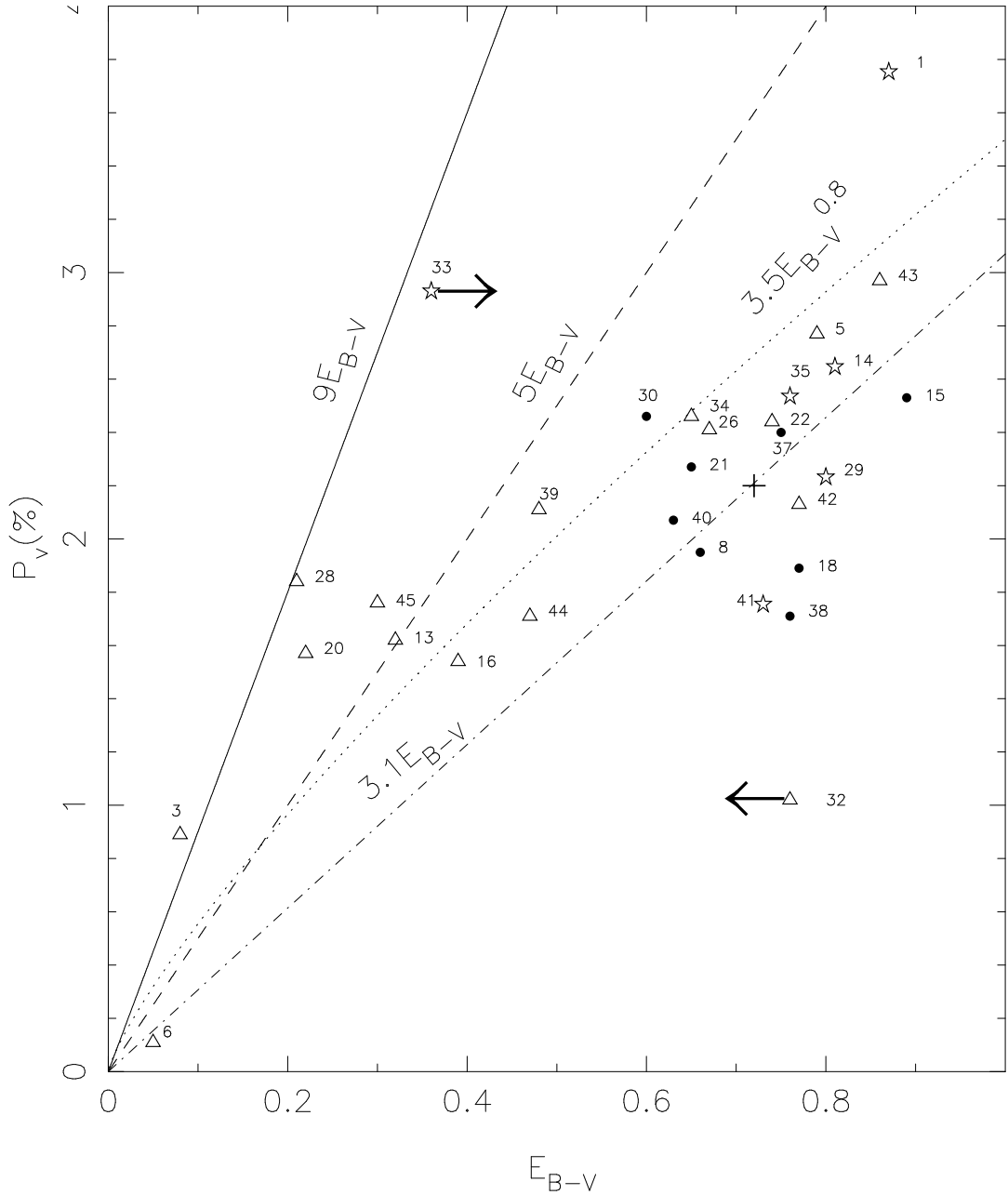}
    \caption{ Plot $P_{V}$ vs. $E_{B-V}$ for stars in the region of NGC 6124. From left to right the
solid line is the maximun polarimetric efficiency  $P_{V} = \ 9 E_{B-V}$, the dashed line is the Serkowki's estimate average for the Galaxy $P_{V} = \ 5 E_{B-V}$, the dotted line is the Fosalba's estimate $P_{V}= 3.5 E(B-V)^{0.8}$, and the dot-dashed line is the efficiency found for the cluster $P_{V}= 3.1 E_{B-V}$. Symbols are as in Fig. 2, but the memberships were modified as discussed in section 4.2.}
\end{figure*}

\begin{figure*}
        \centering
        \includegraphics[width=15cm,angle=-90]{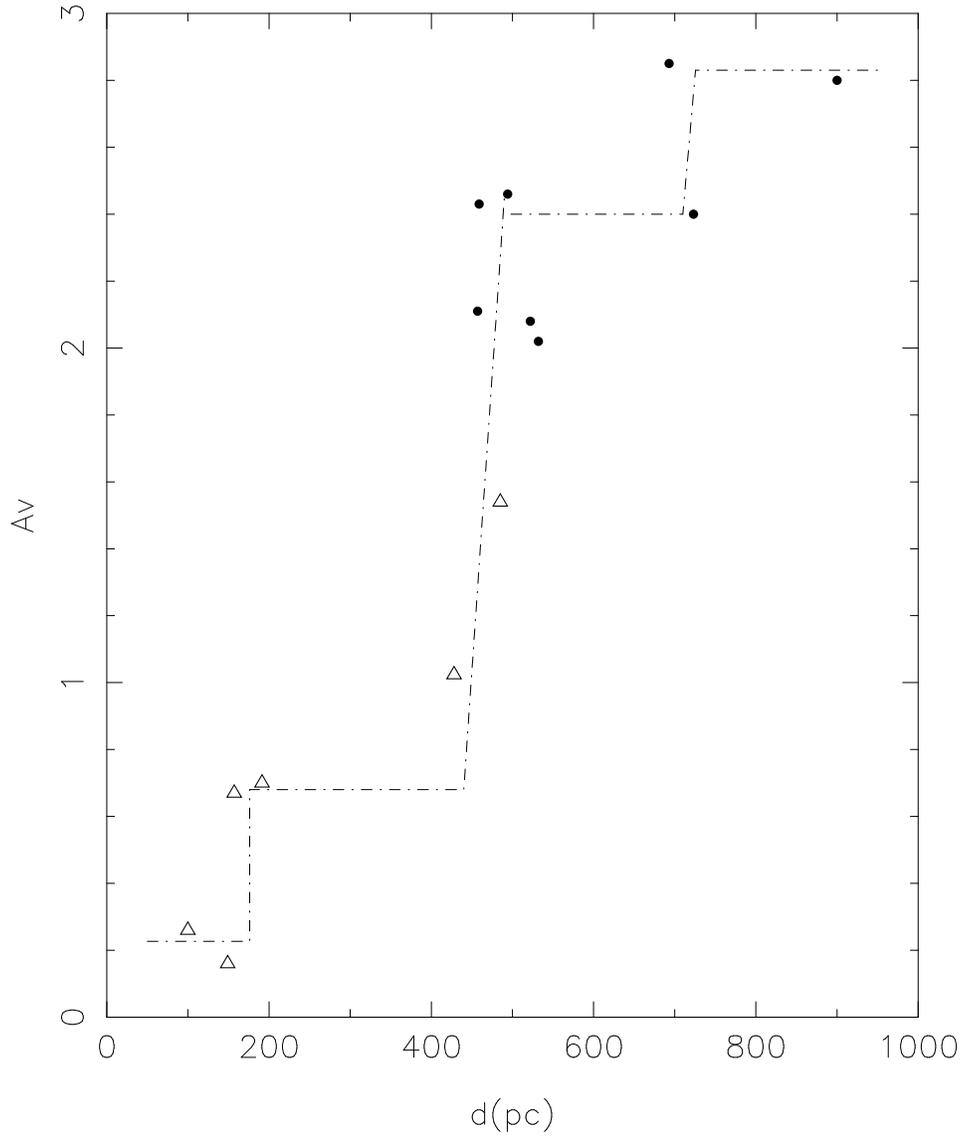}
         \caption{ Variation of the absorption (${A}_V$) with increasing
distance in the area of NGC 6124. This figure includes only those stars whose distances could be
estimated. Open triangles are non-members and solid circles are members.}
\end{figure*}

\end{document}